# Single Image Super-Resolution

Baran Ataman[1], Mert Seker[1], David Mckee[1]

Tampere University, Tampere, Finland

**Abstract**

*This study presents a chronological overview of the single image super-resolution problem. We first define the problem thoroughly and mention some of the serious challenges. Then the problem formulation and the performance metrics are defined. We give an overview of the previous methods relying on reconstruction based solutions and then continue with the deep learning approaches. We pick 3 landmark architectures and present their results quantitatively. We see that the latest proposed network gives favorable output compared to the previous methods.*

.

## 0. Authors' Contributions

Baran Ataman was responsible for Residual Dense Network (RDN) implementation.
Mert Seker was responsible for Super-Resolution Generative Adversarial Network (SRGAN) implementation.
David Mckee was responsible for Super-Resolution Convolutional Neural Network (SRCNN) implementation.
The Report was written by Baran Ataman.

## 1. Introduction

Single image Super-Resolution (SISR) aims to generate visually pleasing high-resolution (HR) images from their degraded low-resolution (LR) versions. It has practical usage in various computer vision tasks, such as security and surveillance imaging, medical image processing, satellite imaging and biometrics recognition. It is an ill-posed inverse problem because under various different circumstances, different HR images can give the same LR images and it is hard to detect the latent HR image among the candidates. To tackle this inverse problem, plenty of image SR algorithms have been proposed, including interpolation-based, reconstruction-based [4, 5], and learning-based methods [2, 3, 7, 8, 9, 10]. The rest of the paper will be discussing some of the significant previous works in section 2, performance metrics in section 3, detailed architectures and implementation details in section 4, description of the data set for our tests in section 5, experiments and results in section 6 and conclusion and discussion in chapter 7.

## 2. Related Work

Interpolation based methods try to recover the latent high-resolution image by interpolating the pixel values with different interpolation functions such as wavelets [11], B-Splines [12] or directional filters [13] as well as multisurface fitting [14]. Although early linear interpolators give weak performance, recent nonlinear approaches produce comparatively more satisfactory results. Among them, Zhang and Wu [14] proposed an edge-guided technique through directional filtering and data fusion. Their method has been the most successful one among the interpolation based methods in terms of quantitative metrics.

Reconstruction based methods try to come up with a forward model to recover the high resolution image. The training of reconstruction based methods need paired data: LR and HR versions of the same image. Since it is not likely to obtain the same scene with same camera in HR and LR setting, the LR image is obtained by down-sampling the HR image and reconstruction based methods try to recover that HR image from its down-sampled LR version. This makes SISR a typical inverse problem which needs a strong forward model to uncloak the latent HR image. Among reconstruction based methods, gradient profile prior [4] has been introduced relying on the fact that the shape statistics of the gradient profiles in natural images is robust against changes in image resolution. This approach tries to learn the similarity between the shape statistics of the LR and HR images. Another important approach was using primal sketches [5] as the a priori. Having an LR input image, they first interpolate it to the target resolution, then for every primitive point (basically every point on the contours inside the image) a small patch is taken. Then, based on the primal sketch prior, the corresponding HR patch for every LR patch is found and replaced. This step hallucinates the high-frequency counterparts of the primitives. Even though several other approaches were developed for SISR in reconstruction framework, the 2 we mentioned were the most remarkable approaches and made the highest jump in the qualitative performance.

Learning based methods can be listed as feature pyramids [15], belief networks [16], projection-based methods [17], manifold learning [18], neural networks and compressive sensing [19]. Recent advances in deep learning produced impressive results in SISR as well as in the other domains of computer vision. First convolutional neural network (CNN) based method proposed by Dong et al. [7] established an end-to-end mapping between the interpolated LR images and their HR counterpart. It was a 3-layer CNN and was improved further by adding more convolutional layers. Another significant approach was SRGAN [3] which utilizes a Generative Adversarial Network (GAN) architecture which employs a deep residual network (ResNet) [20] with skip-connection and diverge from MSE as the sole optimization target. Different from previous works, they define a novel perceptual loss using high-level feature maps of the VGG network. Their discriminator network is a 7-convolutional layer network which is used to distinguish real samples and fake samples. The latest state-of-the-art method is proposed by Zhang et al. named Residual Dense Network (RDN) [2] which employs dense connections and residual learning at the same time. Their objective was to make use of all the hierarchical features of the low resolution image. Skip connection is the widespread resolution to vanishing gradient problem and an effective way of propagating features in the deep networks. [23] was the milestone for adopting dense skip connections which resulted in favorable performance against the previous methods. [22] proposed the enhanced version of SRResNet [21] which removes redundant modules from SRResNet as well as employing residual scaling techniques to achieve stability in the training process.

Our research has its roots on these 3 deep learning methods: the first CNN based approach SRCNN, the first and landmark GAN approach SRGAN and the recent and most successful approach RDN. We will evaluate the performance of these 3 methods on the same data set in order to have consistent evaluation. Next section presents the 2 main performance metrics used in our research, Peak Signal-to-Noise Ratio (PSNR) and Structural Similarity Index (SSIM).

## 3. Performance Metrics
### 3.1 PSNR

PSNR is the ratio between the maximum possible power of a signal and the power of noise that affects the accuracy of its representation. PSNR is usually expressed in terms of the logarithmic decibel scale because many signals have a very wide dynamic range: in 8x3 bit images we have 256 different values. The formula for PSNR is:

$$PSNR = 10\log(\frac{MAX\ Signal\ Value^2}{Total\ Mean\ Squared\ Error})$$

Basically what it does is calculating the mean squared error of all the pixels and using this mean squared error value in the above formula. PSNR is the widely used metric for super-resolution, denoising, deblurring and various other computer vision tasks. It looks at the pixel-wise intensity differences.

### 3.2 SSIM

Another popular metric is the SSIM [24] which is more sophisticated while comparing two images. SSIM is based on visible structures in the image unlike PSNR and it is widely used in compression tasks. The formula for SSIM is below:

$$SSIM(x,y) = \frac{(2\mu_x\mu_y + c1)(2\sigma_{xy} + c2)}{(\mu_x^2 + \mu_y^2 + c1)(\sigma_x^2 + \sigma_x^2 + c2)}$$

where

$\mu_x$ is average of x
$\mu_y$ is average of y
$\sigma_x$ is variance of x
$\sigma_y$ is variance of y
c1 and c2 are stabilization factor

SSIM is started to be used more frequently as the primary indicator in super-resolution and compressions tasks because it measures the perceptual difference between two images.

## 4. Methods
### 4.1 RCNN

The proposed architecture of SRCNN is below:

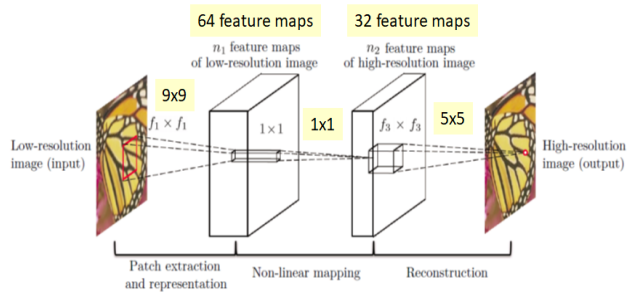

**Figure 1 Architecture of RCNN**

As can be seen above, it is relatively a simple network with 3 parts: patch extraction and representation, non-linear mapping, and reconstruction. The LR input is first upscaled to the desired size using bicubic interpolation and fed into the network. The loss function is the average of mean squared error (MSE) for the training samples:

$$L(\Theta) = \frac{1}{n}\sum_{i=1}^{n}\|F(Y_i;\Theta) - X_i\|^2$$

### 4.2 SRGAN

The architecture of SRGAN is provided below:

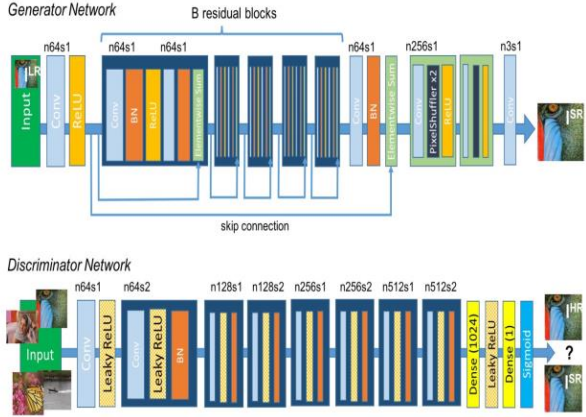

**Figure 2 Architecture of SRGAN**

The generator network is a novel CNN developed by the authors which is called Super-Resolution Residual Network (SRResNet) which is influenced by ResNet and has skip-connections and differ from MSE as the loss function. They defined a novel perceptual loss using high-level feature maps of the VGG network. They produce HR results with this SRResNet and give those results alongside with real HR images to the discriminator network in order to enhance the performance of the method. The loss function the utilized is:

$$L = L_{content\ loss} + 10^{-3} L_{adversarial\ loss}$$

where

$$L_{adversarial\ loss} = \sum_{n=1}^{N} -\log D_{\theta_D}\left(G_{\theta_G}(I^{LR})\right)$$

Content loss can be calculated as the conventional MSE but the authors proposed the VGG loss which measures the MSE of features extracted by a VGG-19 network. The formula for content loss is below:

$$l_{VGG/i.j}^{SR} = \frac{1}{W_{i,j}H_{i,j}}\sum_{x=1}^{W_{i,j}}\sum_{y=1}^{H_{i,j}}(\phi_{i,j}(I^{HR})_{x,y} - \phi_{i,j}(G_{\theta_G}(I^{LR}))_{x,y})^2$$

$\phi_{i,j}$ The feature map for the j-th convolution (after activation) before the i-th maxpooling layer.

### 4.3 RDN

RDN is proposed in 2018 and its main aim is to fully make use of all the hierarchical features from the original LR image with residual dense block. The architecture of the network has 3 main parts: shallow feature extraction, Residual Dense Blocks (RDBs) and dense feature fusion. The architecture of the whole network is below:

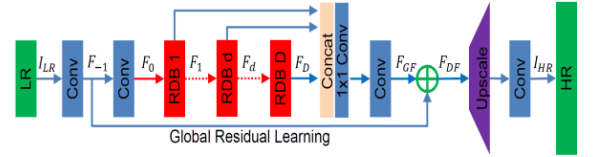

**Figure 4 Architecture of RDN**

As can be seen above, the building blocks of this network are the RDBs and their architecture is below:

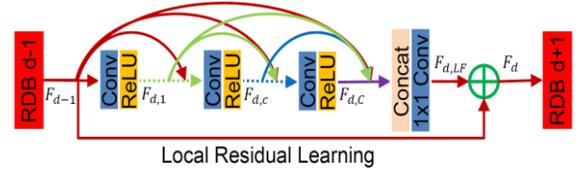

**Figure 5 Architecture of Residual Dense Blocks**

RDN has 3 significant properties:
- The network makes full use of all the hierarchical features from the original LR image.
- RDBs read state from the preceding RDB via a contiguous memory (CM) mechanism, and also fully utilize all the layers within it via local dense connections. The accumulated features are then adaptively preserved by local feature fusion (LFF).
- They implement global feature fusion to adaptively fuse hierarchical features from all RDBs in the LR space. With global residual learning, they combine the shallow features and deep features, resulting in global dense features from the original LR image.

## 5. Dataset

Recently, Timofte et al. have released a high-quality (2K resolution) dataset DIV2K [25] for image restoration applications. DIV2K consists of 800 training images, 100 validation images, and 100 test images. It is originally introduced for NTIRE 2017 Challenge on Single Image Super-Resolution at CVPR 2017. It has become a significant benchmark dataset for super-resolution tasks.

## 6. Experiments and Results

We have downloaded the pre-trained models from Github repositories of the authors and tested them on the DIV2K dataset. We applied x4 high resolution in our experiments. We used the validation set of the data set since test set is not shared publicly. Collecting 100 HR results from 3 different deep learning approaches and the ground truth images, we can present our results:

| Methods/Performance | PSNR | SSIM |
| --- | --- | --- |
| **SRCNN** | 27.23 | 0.76 |
| **SRGAN** | 31.53 | 0.75 |
| **RDN** | 32.47 | 0.80 |

**Table 1 Results of 3 methods in PSNR and SSIM**

As we can see from the above table, RDN outperforms the other two methods in our quantitative metrics. SRGAN beats SRCNN in PSNR with a huge gap but performs almost the same in SSIM metric. We should provide some visual examples to further illustrate the results.

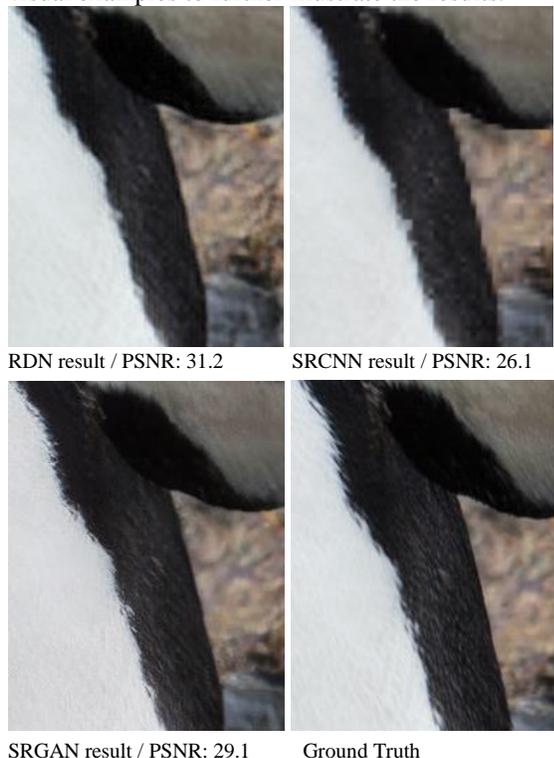

RDN result / PSNR: 31.2    SRCNN result / PSNR: 26.1

SRGAN result / PSNR: 29.1    Ground Truth

**Figure 6 Illustrative figure comparing the PSNR values of the methods**

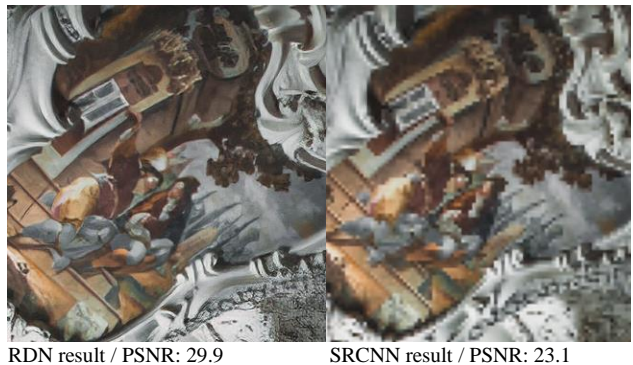

RDN result / PSNR: 29.9    SRCNN result / PSNR: 23.1

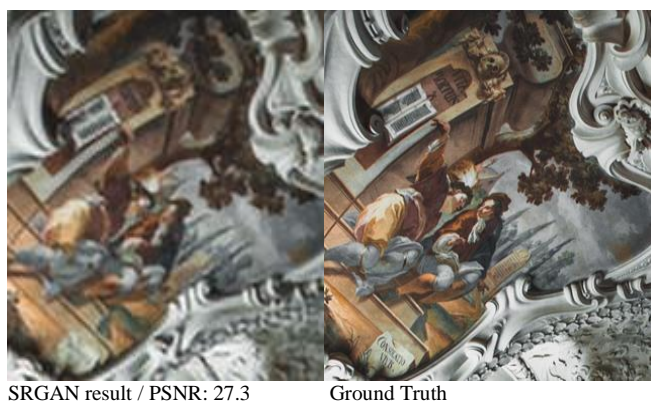

SRGAN result / PSNR: 27.3    Ground Truth

**Figure 7 Illustrative figure comparing the PSNR values of the methods**

As can be seen from the above illustrations, the RDN surpassed the other methods in both quantitative and qualitative metrics: we can safely say the RDN produced visually favorable responses compared to the other methods.

## 7. Conclusion and Discussion

In this paper we presented the 3 remarkable deep learning approaches and compared their performances. Our selected methods were the first CNN based approach, first GAN based approach and the recent state-of-the-art approach. Our experiments were fair because we used the same dataset and applied the same high-resolve ratio to the test images. We saw that the recent proposed RDN network gives the best results both qualitatively and quantitatively. We also examined that PSNR is not a reliable metric to make a comparison between the paired images since low PSNR output may be a bit more favorable or includes more details. For instance in Figure 6, we see that highest PSNR is produced with RDN but one can favour SRGAN result because it is less smoother and has more deviations especially in the colour transitions areas. One can prefer a sharper image rather than a smother image. We saw that SRCNN gives the less favorable results in every image because it is the least advanced and sophisticated architecture.

The performance of these methods can be further improved by training the models with new data in order to both enhance their performance and increase their generalizability. Since these networks are trained on different data sets, we could train every one of them with the other data sets so that all 3 will be stronger and will have a more competitive challenge. Single Image Super-Resolution has been a popular topic for decades and the concentration is now on CNN based methods because GANs are instable and their training is unsupervised whereas CNN based methods use paired data and supervised learning, which is more stable. There are 24 accepted papers to IEEE International Conference on Computer Vision and Pattern Recognition (CVPR) 2019 about image super-resolution and 3 of them claim that they have beaten RDN. This illustrates how much attention is paid on super resolution problem.